\documentclass[12pt,preprint]{aastex}

\usepackage{natbib}

\slugcomment{Submitted to The Astrophysical Journal.}

\begin{document}

\title{Gravitational Radiation from an Accreting Millisecond Pulsar
 with a Magnetically Confined Mountain}


\author{A. Melatos and D. J. B. Payne}
\affil{School of Physics, University of Melbourne,
 Parkville, VIC 3010, Australia}
\email{a.melatos@physics.unimelb.edu.au}


\begin{abstract}
The amplitude of the gravitational radiation from an accreting
neutron star undergoing polar magnetic burial is calculated.
During accretion, the magnetic field of a neutron star 
is compressed into a narrow belt at the magnetic equator 
by material spreading equatorward from the polar cap.
In turn, the compressed field confines the accreted
material in a polar mountain which is misaligned 
with the rotation axis in general,
producing gravitational waves.
The equilibrium hydromagnetic structure of the polar mountain,
and its associated mass quadrupole moment,
are computed as functions of the accreted mass, $M_{\rm a}$,
by solving a Grad-Shafranov boundary value problem.
The orientation- and polarization-averaged gravitational wave strain
at Earth is found to be
$h_{\rm c}=
 6\times 10^{-24}(M_{\rm a}/M_{\rm c})(1+M_{\rm a}b^2/8M_{\rm c})^{-1}
 (f/0.6\,{\rm kHz})^2 (d/1\,{\rm kpc})^{-1}$,
where $f$ is the wave frequency,
$d$ is the distance to the source,
$b$ is the ratio of the hemispheric to polar magnetic flux,
and the cut-off mass $M_{\rm c} \sim 10^{-5} M_{\sun}$ 
is a function of the natal magnetic field,
temperature, and electrical conductivity of the crust.
This value of $h_{\rm c}$ exceeds previous estimates that failed 
to treat equatorward spreading and flux freezing
self-consistently.
It is concluded that an accreting millisecond pulsar emits
a persistent, sinusoidal gravitational wave signal
at levels detectable, in principle, by long baseline interferometers 
after phase-coherent integration,
provided that the polar mountain is hydromagnetically stable.
Magnetic burial also reduces
the magnetic dipole moment $\mu$ monotonically as
$\mu\propto (1+3 M_{\rm a}/4 M_{\rm c})^{-1}$,
implying a novel, observationally testable scaling $h_{\rm c}(\mu)$.
The implications for the rotational evolution of
(accreting) X-ray and (isolated)
radio millisecond pulsars are explored.
\end{abstract}



\newpage 


\section{Introduction
 \label{sec:gra1}}
The current generations of resonant bar antennas and 
long baseline interferometers are capable of detecting 
gravitational wave signals from incoherent sources 
with strain amplitudes $h$ exceeding
$\sim 10^{-20}$ and $\sim 10^{-21}$ respectively
at frequencies near $f\approx 0.6\,{\rm kHz}$
\citep{sch99}.
The sensitivity can be improved by a factor
$(f\tau)^{1/2}$ for periodic sources of known $f$
by integrating coherently for a total observing time $\tau$,
as in heirarchical Fourier searches
\citep{dhu96,bra98,bra00}.
Several classes of sources with promising event rates 
have been identified in the kilohertz regime:
coalescing neutron-star binaries
\citep{phi91},
r-modes in young, hot neutron stars
\citep{and98,lin98},
and neutron stars in low-mass X-ray binaries whose
crusts are deformed by temperature gradients
\citep{bil98,ush00}.
More generally, every isolated and accreting millisecond pulsar
is potentially a kilohertz source, 
because the stellar magnetic field, 
which is usually not symmetric about the rotation axis,
deforms the crust and interior hydromagnetically.
It is presently believed that hydromagnetic
deformations are too small to produce gravitational waves 
detectable by the current generation of interferometers,
with
$h \sim 10^{-31} 
 (B/10^{12}\,{\rm G})^2 (f/0.6\,{\rm kHz})^2
 (d/1\,{\rm kpc})^{-1}$
for an object with surface magnetic field $B$,
situated at a distance $d$ from Earth
\citep{kat89,bon96}.
Magnetars, with $B\gtrsim 10^{15}\,{\rm G}$,
are invoked as a possible exception
\citep{kon00,iok01,pal01},
as are objects whose internal toroidal fields are many times 
greater than $B$
\citep{cut02},
or whose macroscopically averaged Maxwell stress tensor 
is enhanced relative to uniform magnetization
because the internal field is concentrated in flux tubes,
e.g.\ in a type II superconductor
\citep{jon75,bon96}.

In this paper, we show that previous calculations materially
underestimate the hydromagnetic deformation of recycled pulsars,
e.g.\ accreting millisecond pulsars such as SAX J1808.4$-$3658
\citep{wij98,cha98,gal02,mar02}.
During accretion, in a process termed {\em magnetic burial},
material spreads equatorward from the polar cap,
compressing the magnetic field into a narrow belt at the
magnetic equator and increasing the field strength locally
while reducing the global dipole moment $\mu$
\citep{mel00,mel01,pay04},
in accord with observations of low- and high-mass 
X-ray binaries and binary radio pulsars
\citep{taa86,van95}.
In turn, the compressed equatorial magnetic field reacts back 
on the accreted material, confining it in a polar mountain
which is misaligned with the rotation axis in general
\citep{mel00,mel01}.
The gravitational ellipticity $\epsilon$ \citep{bra98}
of the mountain, calculated here self-consistently, 
can approach $\epsilon\sim 10^{-7}$,
materially greater than for an undistorted dipole
[$\epsilon\approx 6\times 10^{-13} (B/10^{12}\,{\rm G})^2$,
e.g.\ \citet{kat89}]
due to the enhanced stress from the compressed field.
A mountain of this size generates gravitational waves at levels
detectable by the current generation of interferometers
and can explain the observed clustering of the spin periods of 
accreting millisecond pulsars through the stalling effect
discovered by \citet{bil98}.

In \S\ref{sec:gra2}, 
we review the physics of magnetic burial and calculate the 
hydromagnetic structure of the polar mountain as a function
of accreted mass $M_{\rm a}$ by solving an
appropriate Grad-Shafranov boundary value problem,
connecting the initial and final states self-consistently,
and treating Ohmic diffusion semiquantitatively
\citep{mel01,pay04}.
In \S\ref{sec:gra3}, 
we predict the orientation- and polarization-averaged 
gravitational wave strain $h_{\rm c}(M_{\rm a})$
and compare it against the sensitivity of the
Laser Interferometer Gravitational-Wave Observatory (LIGO).
We also combine $h_{\rm c}(M_{\rm a})$ with $\mu(M_{\rm a})$
from previous work \citep{pay04} to deduce a novel,
observationally testable scaling $h_{\rm c}(\mu)$ 
for accreting millisecond pulsars
\citep{mel00}.
In \S\ref{sec:gra4}, we discuss the implications of magnetic
burial for the sign of the net torque 
acting on an X-ray millisecond pulsar
and the evolution of $h_{\rm c}$ and $\mu$ 
after accretion ends
(e.g.\ in radio millisecond pulsars).
Our conclusions rely on the assumption that the polar mountain
is not disrupted by hydromagnetic (e.g.\ Parker) instabilities;
the justification of this assumption is postponed to future work.
Deformation by magnetic burial also occurs in white dwarfs
\citep{kat89,hey00,cum02}.

\section{Polar magnetic burial
 \label{sec:gra2}}

\subsection{Equilibrium hydromagnetic structure of the polar mountain
 \label{sec:gra2a}}
The equilibrium mass density $\rho({\bf x})$
and magnetic field intensity ${\bf B}({\bf x})$ 
in the polar mountain are determined by
the equation of hydromagnetic force balance,
\begin{equation}
 \nabla p + \rho\nabla\phi - 
 (4\pi)^{-1} (\nabla\times{\bf B})\times{\bf B} = 0~,
\label{eq:gra1}
\end{equation}
supplemented by 
(i) the condition $\nabla\cdot{\bf B}=0$,
(ii) the gravitational potential $\phi$
(with $|\nabla\phi|\approx GM_\ast /R_\ast^2$, 
where $M_\ast$ and $R_\ast$ are the stellar mass and radius,
because the mountain is much shorter than $R_\ast$),
and (iii) an equation of state for the pressure $p$
(isothermal for simplicity, i.e.\
$p=c_{\rm s}^2 \rho$, where $c_{\rm s}$ is the sound speed).
We define spherical polar coordinates $(r,\theta,\phi)$,
such that $\theta=0$ coincides with the magnetic axis
before accretion, and we seek solutions to (\ref{eq:gra1}) 
that are symmetric about this axis,
such that ${\bf B}$ can be constructed from a scalar
flux function $\psi(r,\theta)$ according to
${\bf B}=(r\sin\theta)^{-1}\nabla\psi \times 
 {\bf \hat e}_\phi$.
Upon resolving (\ref{eq:gra1}) into components
parallel and perpendicular to ${\bf B}$,
we arrive respectively at the barometric formula
$\rho = c_{\rm s}^{-2} F(\psi) 
 \exp[-(\phi-\phi_0)/c_{\rm s}^2]$,
with $\phi_0=GM_\ast/R_\ast$,
and the Grad-Shafranov equation describing cross-field force balance
\citep{ham83,bro98,lit01,mel01,pay04},
\begin{equation}
  \frac{\partial^2\psi}{\partial r^2} +
  \frac{\sin\theta}{r^2} \frac{\partial}{\partial\theta}
  \left(
   \frac{1}{\sin\theta} \frac{\partial\psi}{\partial\theta}  
  \right)
 =
 - 4\pi r^2 \sin^2\theta
 F'(\psi) \exp[-(\phi-\phi_0)/c_{\rm s}^2]~.
\label{eq:gra2}
\end{equation}

It is customary to choose an arbitrary (albeit physically plausible)
functional form of $F(\psi)$ in order to solve (\ref{eq:gra2})
\citep{uch81,ham83,bro98,lit01}.
However, this approach leads to an inconsistency.
In the perfectly conducting limit,
material is frozen to magnetic field lines.
Hence the mass
$dM = 2\pi d\psi \int ds\, \rho[r(s),\theta(s)]
 r\sin\theta |\nabla\psi|^{-1}$
enclosed between two adjacent flux surfaces $\psi$ and $\psi+d\psi$
(where the integral is along a field line)
equals the mass enclosed before accretion
plus any mass added during accretion,
{\em without any cross-field transport of material}
\citep{mou74,mel01,pay04}.
To calculate $\psi$ correctly, one must specify $dM/d\psi$
according to the global accretion physics,
with $F(\psi)$ following from
\begin{equation}
 F(\psi) = 
 \frac{c_{\rm s}^2}{2\pi} \frac{dM}{d\psi}
 \left\{
  \int ds\,r\sin\theta |\nabla\psi|^{-1}
  \exp[-(\phi-\phi_0)/c_{\rm s}^2]
 \right\}^{-1} ;
\label{eq:gra4}
\end{equation}
otherwise, if $F(\psi)$ is specified,
$dM/d\psi$ changes as a function of $M_{\rm a}$ in a manner
that inconsistently leads to cross-field transport.
The functional form of $F(\psi)$,
determined self-consistently through (\ref{eq:gra4}),
changes as $M_{\rm a}$ increases.
This property, 
newly recognized in the context of magnetic burial
\citep{mel01,pay04},
has an important astrophysical consequence:
it produces a greater hydromagnetic deformation
than predicted by previous authors,
because polar (accreting) and equatorial (nonaccreting) flux tubes 
maintain strictly separate identities through (\ref{eq:gra4}),
without exchanging material,
and hence the equatorial magnetic field is highly compressed.

We solve (\ref{eq:gra2}) and (\ref{eq:gra4}) simultaneously
for $\psi(r,\theta)$ and $\rho(r,\theta)$ subject to
the line-tying (Dirichlet) boundary condition
$\psi(R_\ast,\theta)=\psi_\ast \sin^2\theta$
at the stellar surface,
such that the footpoints of magnetic field lines
are anchored to the heavy, highly conducting crust
\citep{mel01,pay04}.
We adopt a mass-flux distribution,
$M(\psi)=\case{1}{2} M_{\rm a}(1-e^{-\psi/\psi_{\rm a}})
 (1-e^{-\psi_\ast/\psi_{\rm a}})^{-1}$,
that embodies the essence of disk accretion,
namely that the accreted mass is
distributed rather evenly within the polar flux tube
$0\leq\psi\leq\psi_{\rm a}$, with minimal leakage
onto equatorial flux surfaces 
$\psi_{\rm a}\leq\psi\leq\psi_\ast$,
where $\psi_\ast=\psi(R_\ast,\pi/2)$ denotes the 
hemispheric flux and $\psi_{\rm a}$ is
the flux surface touching the inner edge of
the accretion disk (radius $R_{\rm a}$).
\citet{pay04} verified that the results do not depend
sensitively on the exact form of $M(\psi)$,
only on the ratio $\psi_{\rm a}/\psi_\ast$.

The boundary value problem (\ref{eq:gra2}) and (\ref{eq:gra4})
with surface line-tying can be solved analytically,
in the small-$M_{\rm a}$ limit,
by Green functions\footnote{The Grad-Shafranov 
operator is not self-adjoint.} after approximating the source term
on the right-hand side of (\ref{eq:gra2}) to sever
the coupling between (\ref{eq:gra2}) and (\ref{eq:gra4})
\citep{pay04}.
It can also be solved numerically by an iterative scheme
that solves the Poisson equation (\ref{eq:gra2})
for a trial source term by successive overrelaxation
then updates $F(\psi)$ from (\ref{eq:gra4}) by
integrating numerically along a set of contours
(including closed and edge-interrupted loops)
\citep{mou74,pay04}.
The numerical scheme is valid for arbitrarily large $M_{\rm a}$,
although convergence deteriorates as $M_{\rm a}$ increases.
Typically, we use a $256\times 256$ grid and
$255$ contours (linearly or logarithmically spaced)
and target an accuracy of $|\Delta\psi/\psi|\leq 10^{-2}$
after $\sim 10^3$ iterations.
We rescale the $r$ and $\theta$ coordinates logarithmically 
in the regions where steep gradients develop,
e.g.\ $\theta\approx \pi/2$
\citep{pay04}.

\subsection{Ohmic diffusion
 \label{sec:gra2b}}
The structure of the polar mountain evolves quasistatically,
over many Alfv\'{e}n times, in response to 
(i) accretion, 
which builds up the mountain against the confining stress
of the compressed equatorial magnetic field,
and (ii) Ohmic diffusion,
which enables the mountain to relax equatorward as
magnetic field lines slip through the resistive fluid.
The competition between accretion and Ohmic diffusion 
has been studied in detail in the context of
neutron stars 
\citep{bro98,lit01,cum01}
and white dwarfs
\citep{cum02}.
In these papers,
steady-state, one-dimensional profiles of
the magnetic field are computed as functions of depth, 
from the ocean down to the outer crust,
and the Ohmic ($t_{\rm d}$) and accretion ($t_{\rm a}$)
time-scales are compared,
under the assumptions
that the magnetic field is flattened parallel to the surface
by polar magnetic burial,
the accreted material is unmagnetized,
and the accretion is spherical;
`the complex problem of the subsequent spreading of matter'
is not tackled 
\citep{cum02}.
The field penetrates the accreted layer
if the accretion rate satisfies 
$\dot{M}_{\rm a} \lesssim 0.1 \dot{M}_{\rm Edd}$,
where $\dot{M}_{\rm Edd}$ is the Eddington rate,
but it is screened diamagnetically (i.e.\ buried)
if $\dot{M}_{\rm a} \gtrsim 0.1 \dot{M}_{\rm Edd}$,
such that the surface field is reduced
$\approx (\dot{M}_{\rm a} / 0.002 \dot{M}_{\rm Edd})$-fold
relative to the base of the crust
\citep{cum01}.\footnote{
SAX J1808.4$-$3658 is presumed to possess an ordered
magnetic field at its surface because it pulsates
\citep{wij98,cha98}.
Due to the low accretion rate,
$\dot{M}_{\rm a}\approx 10^{-11} M_{\sun}\,{\rm yr^{-1}}$,
either the field has penetrated to the surface by Ohmic diffusion
\citep{cum01},
or polar magnetic burial and equatorward spreading
have not proceeded far enough
\citep{pay04}.
}

In the regime $t_{\rm d}\lesssim t_{\rm a}$,
Ohmic diffusion outpaces accretion.
As material is added, it does not compress
the equatorial magnetic field further;
instead, it diffuses across field gradients and
distributes itself uniformly over the stellar surface. 
Hence the polar mountain
(i.e.\ the asymmetric component of $\rho$)
`stagnates' at the structure attained when 
$t_{\rm d}\sim t_{\rm a}$.
The accretion time-scale is defined as 
$t_{\rm a}=M_{\rm a}/\dot{M}_{\rm a}$.
The Ohmic diffusion time-scale is given by
$t_{\rm d} = 4\pi \sigma L^2/c^2$,
where $\sigma$ denotes the electrical conductivity
and $L=(|\psi|/|\nabla\psi|)_{\rm min}$ is the 
characteristic length-scale of the steepest field gradients;
$L$ reduces to the hydrostatic scale-height in the
one-dimensional geometry employed in earlier work
\citep{bro98,cum01,cum02}
but is dominated by latitudinal gradients here.
Following \citet{cum01},
we assume that the electrical resistivity is dominated
by electron-phonon scattering in the outer crust
($M_{\rm a} \gtrsim 10^{-10} M_{\sun}$),
as expected if the crustal composition is primordial,
although electron-impurity scattering 
may dominate if the products of hydrogen/helium burning 
leach into the crust.
In the relaxation time approximation,
with all Coulomb logarithms set to unity,
the electron-phonon conductivity is given by
$\sigma=7.6\times 10^{26}
 (\rho/10^{11}\,{\rm g\,cm^{-3}})
 (\nu_{\rm c}/10^{16}\,{\rm s^{-1}})^{-1}
 (m_{\rm eff}/m_{\rm e})^{-1} (A/2)^{-1}
 \,{\rm s^{-1}}$,
where $A$ and $Z$ denote the mean molecular weight 
and atomic number per electron,
$m_{\rm eff}$ is the effective electron mass,
$\nu_{\rm c} = 1.2\times 10^{18}
 (T/10^8\,{\rm K})\,{\rm s^{-1}}$
is the electron-phonon collision frequency,
$T$ is the temperature of the crust,
and
$\rho= 6.2\times 10^{10}
 A Z^{-1} (M_{\rm a}/10^{-5} M_{\sun})^{3/4}
 \,{\rm g\,cm^{-3}}$
is the density at the base of the accreted layer
\citep{bro98,cum01}.
Ohmic diffusion therefore arrests the growth of the
polar mountain for
\begin{equation}
 \frac{|\psi|}{|\nabla\psi|}
 \lesssim 
 4.2\times 10^2 Z^{1/2} 
 \left( \frac{M_{\rm a}}{10^{-5} M_{\sun}} \right)^{1/8}
 \left( \frac{T}{10^8\,{\rm K}} \right)^{1/2}
 \left( \frac{m_{\rm eff}}{m_{\rm e}} \right)^{1/2}
 \left( \frac{\dot{M}_{\rm a}}{\dot{M}_{\rm Edd}} \right)^{-1/2} 
 {\rm cm}~,
\label{eq:gra5}
\end{equation}
corresponding to the condition $t_{\rm d}\leq t_{\rm a}$.
The left-hand side of (\ref{eq:gra5}) is computed directly from 
the numerical solution, by scanning over the grid, or from the
approximate analytic solution in \S\ref{sec:gra2c};
it depends implicitly on $M_{\rm a}$.
We denote by $M_{\rm d}$ the minimum accreted mass
for which (\ref{eq:gra5}) is satisfied.
Note that $t_{\rm d}/t_{\rm a}$ is constant with depth
for electron-phonon scattering but decreases with depth
for electron-impurity scattering
\citep{bro98,cum01}.

Several second-order effects, neglected in (\ref{eq:gra5}),
are postponed to future work.
The Hall conductivity vanishes in plane-parallel geometry 
\citep{cum01}
but more generally amounts to a fraction
$1.8\times 10^{-3} (B/10^8\,{\rm G})
 (\nu_{\rm c} / 10^{18}\,{\rm s^{-1}})^{-1} 
 (m_{\rm eff}/m_{\rm e})^{-1}$
of $\sigma$
\citep{cum01,cum04},
i.e.\ the electron cyclotron frequency divided by $\nu_{\rm c}$.
It may therefore dominate near the equator, 
where the compressed field can reach $B\sim 10^{15}\,{\rm G}$.
In addition, the process of thermomagnetic drift
is negligible in the ocean and outer crust,
but it can dominate in the thin hydrogen/helium skin 
overlying the ocean
\citep{gep94,cum01}.
Neither the Hall nor the thermomagnetic drifts are diffusive;
they do not smooth out field gradients as $\sigma$ does.
Indeed, the Hall drift tends to intensify field gradients by
twisting the magnetic field in regions where the velocity
of the electron fluid is sheared.
The time-scale of this process is sensitive to the field geometry
(as well as the radial profiles of the electron density
and elastic shear modulus in the crust)
\citep{cum04};
it will be interesting to evaluate it for the distorted field
produced by magnetic burial.

\subsection{Mass quadrupole and magnetic dipole moments
 \label{sec:gra2c}}
\begin{figure}
\plottwo{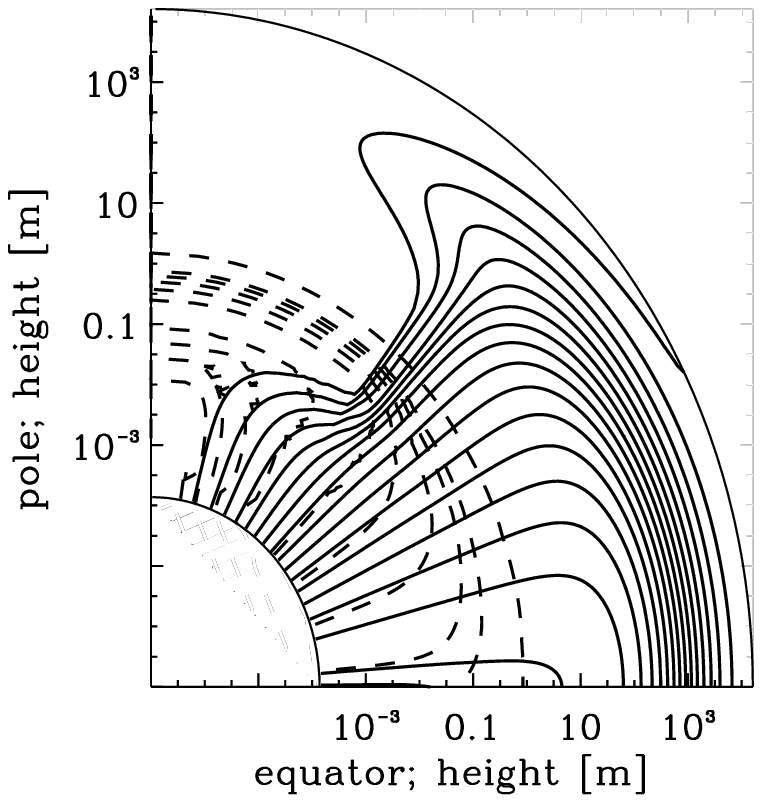}{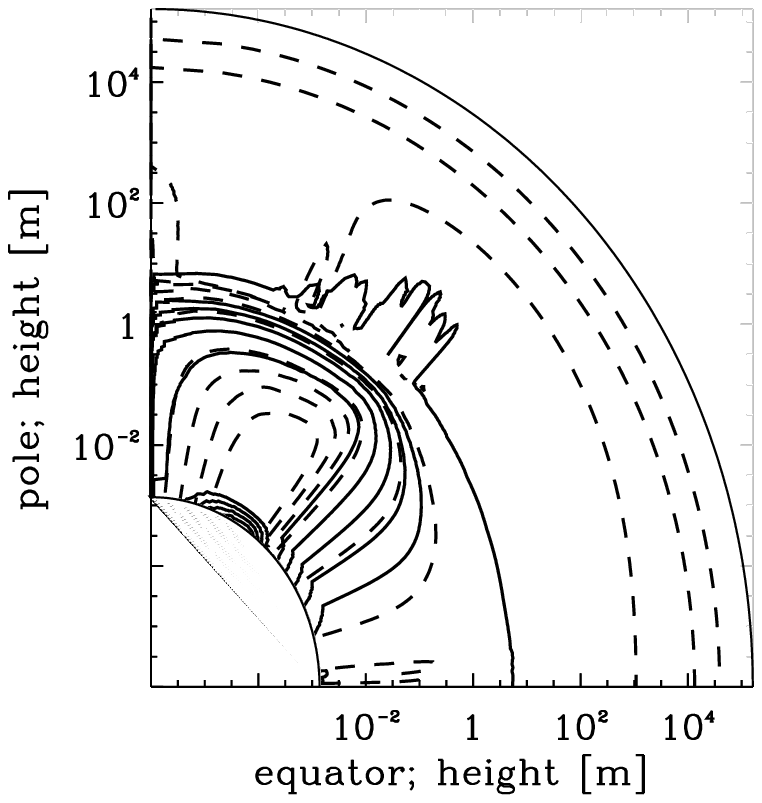}
\caption{
($a$) Numerical calculation of the hydromagnetic structure 
of the accreted layer, showing contours of $\psi$ (solid) 
and $\rho$ (dashed).
($b$) Contours of Lorentz force density (solid) and
magnetic field intensity (dashed).
Parameters: $M_{\rm a}=10^{-5} M_{\sun}$, $b=10$.
The contours are at fractions
$\eta=0.8$, 0.6, 0.4, 0.2, $10^{-2}$, $10^{-3}$,
$10^{-4}$, $10^{-5}$, $10^{-6}$, and $10^{-12}$
of the respective maxima.
\label{fig:gra1}}
\end{figure}

Figure \ref{fig:gra1}$a$ depicts the density profile
of the polar mountain (dashed $\rho$ contours) and 
the flaring geometry of the magnetic field 
(solid $\psi$ contours),
for $M_{\rm a}=10^{-5} M_{\sun}$ and 
$\psi_{\rm a}/\psi_\ast = 10^{-1}$.
The accreted material spreads equatorward under its
own weight until its advance is halted near the equator 
by the stress of the compressed (and hence amplified,
by flux conservation) magnetic field.
Figure \ref{fig:gra1}$b$ illustrates this balance of forces;
the Lorentz force per unit volume,
$(4\pi)^{-1} (\nabla\times{\bf B})\times{\bf B}$
(solid contours),
peaks at the boundary $\psi_{\rm a}$ of
the polar flux tube that receives accreted material,
while the local magnetic field intensity,
$|{\bf B}|$ (dashed contours),
is amplified $\approx 4\times 10^3$ times above
its initial polar value.

\begin{figure}
\plottwo{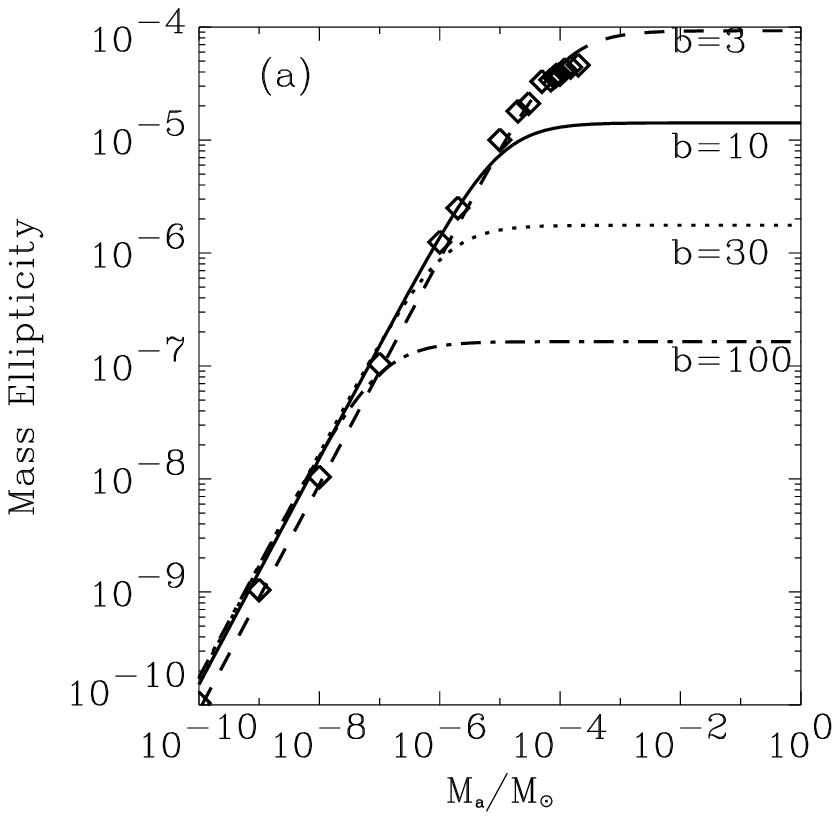}{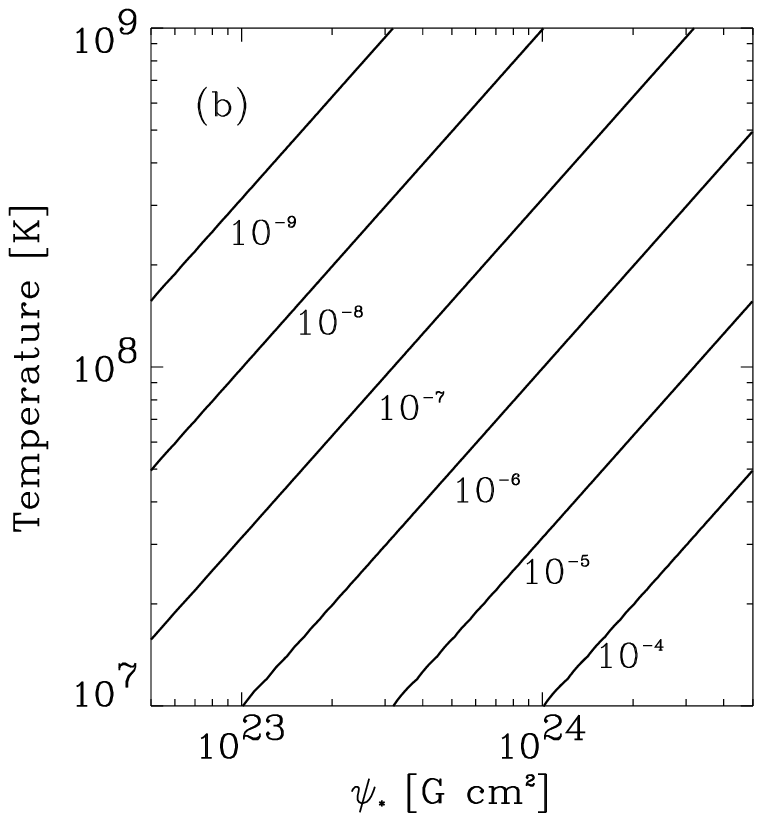}
\caption{
($a$) Ellipticity $\epsilon=|I_1-I_3|/I_1$ as a function of 
accreted mass $M_{\rm a}$, 
for $b=\psi_\ast/\psi_{\rm a}=3$, $10$, $30$, $100$.
The curves are theoretical, based on the expressions
(\ref{eq:gra6a})--(\ref{eq:gra6e}).
The points are numerical results for $b=3$.
Parameters: $\psi_\ast=10^{24}\,{\rm G\,cm^2}$, $T=10^8\,{\rm K}$.
($b$)
Contours of maximum ellipticity 
$\epsilon_{\rm max} =20 M_{\rm c}/(M_\ast b^2)$,
from (\ref{eq:gra6c}) and (\ref{eq:gra6e}),
as a function of total magnetic flux $\psi_\ast$ and
crustal temperature $T$. 
Contours are spaced logarithmically in the range
$10^{-9} \leq M_{\rm c}/M_{\sun} \leq 10^{-4}$.
In both panels, the regime $t_{\rm d} > t_{\rm a}$
is considered to highlight the effects of burial in isolation.
\label{fig:gra2}}
\end{figure}

In Figure \ref{fig:gra2}$a$, we plot the gravitational ellipticity
$\epsilon=|I_1-I_3|/I_1$,
where $I_1$ and $I_3$ denote principal moments of inertia,
as a function of $M_{\rm a}$
for several values of $\psi_{\rm a}/\psi_{\ast}$.
The reduced mass quadrupole moment is then given by
$2 \epsilon I_{zz} / 3$,
with $I_{zz} \approx 0.4 M_\ast R_\ast^2$
\citep{sha83,bon96}.
There is good agreement between the
numerical and analytic results for
$\psi_{\rm a}/\psi_\ast = 0.3$,
where the gradients are manageable and the code converges reliably.
The analytic results follow from a Green function analysis
in the small-$M_{\rm a}$ limit, where the source term
on the right-hand side of (\ref{eq:gra2}) is evaluated
for a dipole, yielding
\citep{pay04}
\begin{equation}
 \psi(x,y)
 =
 \psi_\ast (1-y^2) (1+x/a)^{-1}
 [ 1 - (M_{\rm a}/M_{\rm c}) b^2 y g(y) f(x) ]~,
\label{eq:gra6a}
\end{equation}
\begin{eqnarray}
 \rho(x,y) 
 & = &
 M_{\rm a} a b [ 2\pi R_\ast^3 f(b) ]^{-1}
 [
  y^2 + (1-y^2) (M_{\rm a}/M_{\rm c}) b^2 y g(y) f(x)
 ]^{1/2}
 \\ \nonumber
 & & \times
 \exp[
  -b (1-y^2) - x + (M_{\rm a}/M_{\rm c}) b^3 y g(y) f(x)
 ]~,
\label{eq:gra6b}
\end{eqnarray}
\begin{equation}
 \mu
 = 
 \psi_\ast R_\ast (1 + 3M_{\rm a}/4M_{\rm c})^{-1} ~,
\label{eq:gra6c}
\end{equation}
\begin{equation}
 \epsilon
 = 
 (5 M_{\rm a}/ 2 M_\ast )
 ( 1 - 3/2b ) (1 + M_{\rm a} b^2 / 8 M_{\rm c} )^{-1} ~,
\label{eq:gra6d}
\end{equation}
to leading order in $M_{\rm a}/M_{\rm c}$, with 
\begin{equation}
 M_{\rm c} = 
  G M_\ast \psi_\ast^2 / 4 c_{\rm s}^4 R_\ast^2~.
\label{eq:gra6e}
\end{equation}
We define the dimensionless quantities
$x= a (r/R_\ast -1)$,
$y=\cos\theta$,
$f(x)=1-\exp(-x)$,
$g(y)=\exp[-b(1-y^2)]$,
$a=GM_\ast/c_{\rm s}^2 R_\ast$,
and
$b=\psi_\ast/\psi_{\rm a}$.
Typically, 
the hydrostatic scale height is small compared to $R_\ast$,
as is the polar fraction of the hemispheric flux,
implying $a\gg 1$ and $b\gg 1$ respectively.
The critical accreted mass $M_{\rm c}$
depends on the natal magnetic flux and crustal temperature
through (\ref{eq:gra6e}),
with 
$M_{\rm c}/M_{\sun} = 3\times 10^{-4}
 (\psi_\ast/10^{24}\,{\rm G\,cm^2})^2 
 (T/10^8\,{\rm K})^{-2}$.

The $\epsilon$($M_{\rm a}$) relation (\ref{eq:gra6d}) 
associated with magnetic burial is new,
while the $\mu$($M_{\rm a}$) relation (\ref{eq:gra6c})
is closely related to the phenomenological scaling
invoked by \citet{shi89}
to model observations of binary and millisecond radio pulsars
\citep{taa86,van95}.
It is important to note that magnetic burial affects
$\epsilon$ and $\mu$ in different ways.
The accreted mass above which $\epsilon$ saturates,
$8 M_{\rm c} b^{-2}$,
and the saturation ellipticity,
$\epsilon_{\rm max} = 20 (M_{\rm c}/M_\ast) b^{-2} (1-3/2b)$,
are inversely proportional to $b^2$ (for $b\gg 1$),
whereas the accreted mass above which $\mu$ is screened,
$4 M_{\rm c}/3$, 
is independent of $b$.
This theoretical prediction
conforms with observations in two key respects.
First,
most accreting millisecond pulsars have $b \gtrsim 10$
\citep{lit01}
and hence 
$\epsilon_{\rm max} \lesssim 10^{-7}$ 
from (\ref{eq:gra6d}),
consistent with the upper limit $\epsilon\lesssim 10^{-5}$ 
inferred from spin down
\citep{dhu96,bra98}
and the failure of bar antennas 
and interferometers to detect gravitational waves so far
\citep{sch99}.
Contours of $\epsilon_{\rm max}$
are plotted in the $\psi_\ast$-$T$ plane 
in Figure \ref{fig:gra2}$b$.
Second, the floor magnetic moment of recycled neutron stars,
which is observed to be `universal',
is given by
$\mu_{\rm min} \sim 10^{26}\,{\rm G\,cm^{-3}}$ 
(for $M_{\rm a}\sim 10^{-1}M_{\sun}$) from (\ref{eq:gra6c}).
Theoretically,
it is independent of $b$ (and hence $\dot{M}_{\rm a}$) 
in the regime $t_{\rm d} \gtrsim t_{\rm a}$,
and weakly dependent on $\dot{M}_{\rm a}$ 
in the regime $t_{\rm d} \lesssim t_{\rm a}$
\citep{shi89,mel01}.

\begin{figure}
\plotone{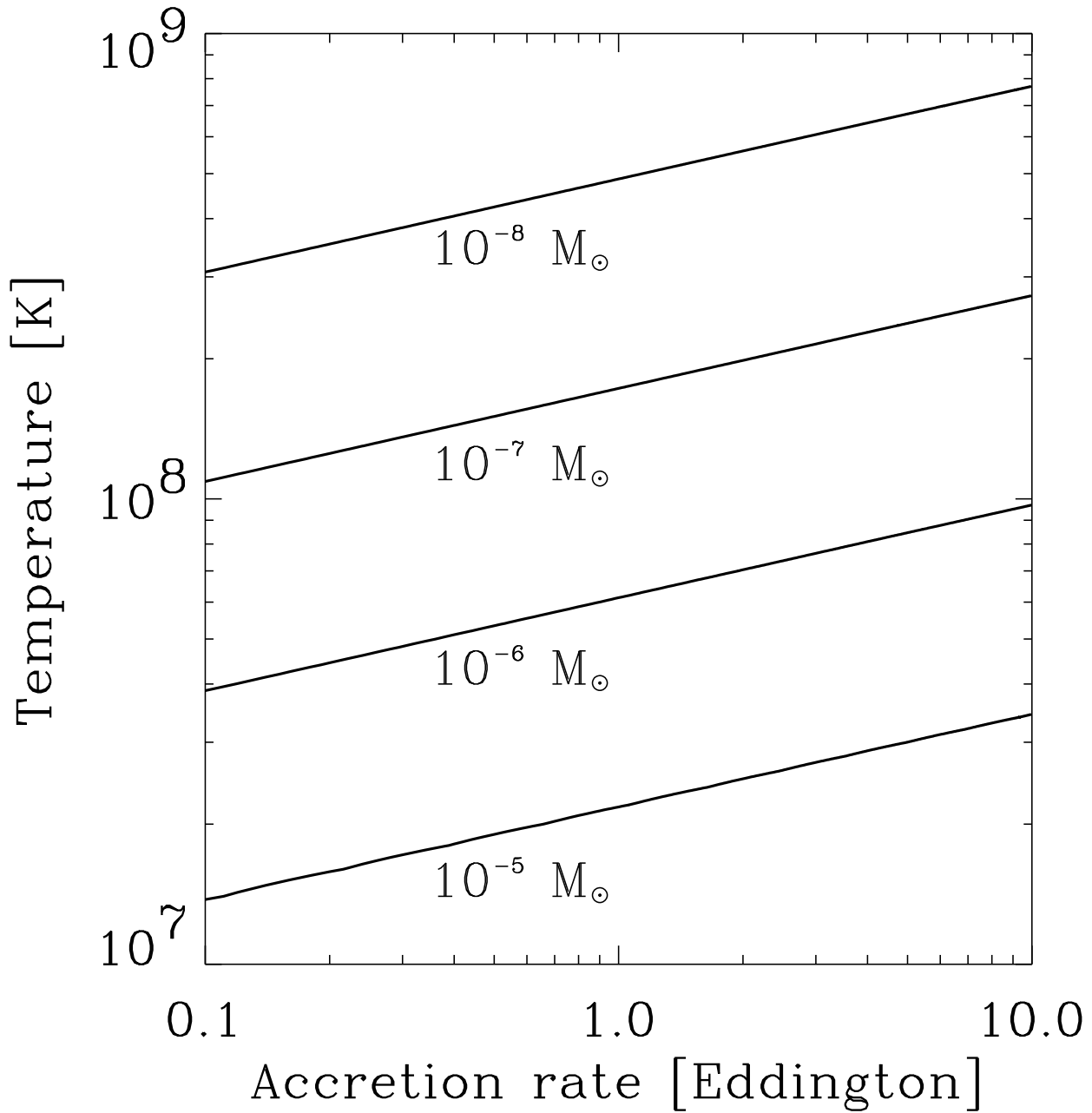}
\caption{
Contours of $M_{\rm d}$, the minimum accreted mass for which
Ohmic diffusion arrests the growth of the polar mountain
($t_{\rm d} \leq t_{\rm a}$). The plot shows the variation
of $M_{\rm d}$ with accretion rate $\dot{M}_{\rm a}$ 
(in units of $\dot{M}_{\rm Edd}$)
and crustal temperature $T$ (in kelvins), in the range
$10^{-8} \leq M_{\rm d}/M_{\sun} \leq 10^{-5}$,
with $\psi_{\rm a}/\psi_\ast = 0.1$, $Z=1$,
and $\psi_\ast = 10^{24}\,{\rm G\,cm^2}$.
$M_{\rm d}$ is calculated from
(\ref{eq:gra5}), (\ref{eq:gra6a}), and (\ref{eq:gra6e}).
\label{fig:gra0}}
\end{figure}

The dipole and quadrupole moments saturate at
$\mu(M_{\rm d})$ and $\epsilon(M_{\rm d})$ respectively
when Ohmic diffusion dominates 
($M_{\rm a} > M_{\rm d}$; see \S\ref{sec:gra2b}).
From (\ref{eq:gra5}), (\ref{eq:gra6a}), and (\ref{eq:gra6e}),
we obtain
\begin{equation}
 \frac{M_{\rm d}}{M_{\sun}}
 = 
 3.4\times 10^{-7} Z^{-4/9} 
 \left( \frac{\psi_{\rm a}}{0.1\psi_\ast} \right)^{16/9}
 \left( \frac{T}{10^8\,{\rm K}} \right)^{-20/9}
 \left( \frac{\psi_\ast}{10^{24}\,{\rm G\,cm^2}} \right)^{16/9}
 \left( \frac{\dot{M}_{\rm a}}{\dot{M}_{\rm Edd}} \right)^{4/9}.
\label{eq:gra5a}
\end{equation}
Contours of $M_{\rm d}$ in the $\dot{M}_{\rm a}$-$T$ plane
are displayed in Figure \ref{fig:gra0}.

The distorted magnetic field in Figure \ref{fig:gra1}$a$ may be
disrupted by interchange, Parker, and doubly diffusive instabilities
wherever the local field strength exceeds 
$B \sim 10^{10}\,{\rm G}$
\citep{cum01,lit01,pay04}.
However, more work is required to settle this issue;
existing stability calculations are linear and plane-parallel,
unlike the situation in Figure \ref{fig:gra1}$a$.
The equilibrium field may not be disrupted completely
if the instability saturates promptly in the nonlinear regime;
for example, the interchange instability is inhibited topologically
by the line-tying boundary condition, which constrains the mobility
of closely packed flux tubes.\footnote{Indeed, the numerical solution 
in Figure \ref{fig:gra1}$a$ represents
the endpoint of a {\em convergent} sequence 
of iterations in a relaxation scheme.
Convergence is cited by \citet{mou74} as
evidence of stability because the relaxation process `mimics'
(albeit imperfectly)
the true time-dependent evolution.}
In a separate effect, \citet{pay04} proved analytically 
that the magnetic field develops bubbles
for $M_{\rm a} \gtrsim 10^{-4} M_{\sun}$;
the source term on the right-hand side of (\ref{eq:gra2})
creates flux surfaces 
$\propto F'(\psi) \propto M_{\rm a}$
that are disconnected from the star ($\psi<0$, $\psi >\psi_\ast$).
Further work is required to determine how the bubbles evolve
in the large-$M_{\rm a}$ regime 
(e.g.\ $M_{\rm a} \sim 10^{-1} M_{\sun}$),
where the theory in \S\ref{sec:gra2a} breaks down.

\section{Gravitational radiation
 \label{sec:gra3}}

\subsection{Detectability 
 \label{sec:gra3a}}
The mass distribution resulting from polar magnetic burial is
not rotationally symmetric in general;
the principal axis of inertia ${\bf e}_3$
(i.e.\ the dipole axis of the natal magnetic field)
is inclined at an angle $\alpha$ to the rotation axis.
Gravitational waves are emitted by the deformed star 
at the spin frequency $\Omega=f/2$ and its first harmonic $2\Omega=f$ 
(unless $\alpha=\pi/2$, when there is no $f/2$ component),
and the spin-down luminosity is proportional to
$(16\sin^2\alpha + \cos^2\alpha) \sin^2\alpha$
\citep{sha83,bon96}.
Upon averaging over $\alpha$, polarization, and orientation
(the position angle of the rotation axis on the sky 
cannot normally be measured),
one can define the characteristic gravitational wave strain
$h_{\rm c} = (128\pi^4 /15)^{1/2} G I_{zz} f^2 \epsilon/ (dc^4)$
\citep{bra98},
which reduces to
\begin{equation}
 h_{\rm c} = 7.7 \times 10^{-19}
 \left( \frac{M_{\rm a}}{M_\ast} \right)
 \left( 1 - \frac{3}{2b} \right)
 \left( 1 + \frac{M_{\rm a} b^2}{8 M_{\rm c}} \right)^{-1}
 \left( \frac{f}{0.6\,{\rm kHz}} \right)^2
 \left( \frac{d}{1\,{\rm kpc}} \right)^{-1}
\label{eq:gra7}
\end{equation}
upon substituting (\ref{eq:gra6d}).
Polar magnetic burial therefore generates 
gravitational radiation whose amplitude
$h_{\rm c} \approx 6\times 10^{-26}$ (for typical parameters
$M_{\rm a}\gtrsim M_{\rm c} \sim 10^{-5} M_\ast$ and $b=30$)
is $\sim 10^5$ times greater than that produced by the natal, 
undistorted magnetic dipole,
$h_{\rm c} \approx 10^{-31} 
 (B/10^{12}\,{\rm G})^2 (f/0.6\,{\rm kHz})^2
 (d/1\,{\rm kpc})^{-1}$
\citep{kat89,bon96},
due to the enhanced Maxwell stress from the
compressed equatorial magnetic field.
The self-consistent form of $F(\psi)$ defined by (\ref{eq:gra4})
is needed to calculate this stress properly;
cf.\ \citet{mel01}.

Figure \ref{fig:gra3}$a$ is a plot of $h_{\rm c}$ versus $f$
for $10^{-8} \leq M_{\rm a}/M_{\sun} \leq 10^{-2}$ and
$10\leq b\leq 10^2$.
The sensitivity curves for initial LIGO ($L$)
and advanced LIGO ($AL$) are superimposed,
corresponding to the weakest source
detectable with 99 per cent confidence in $\tau=10^7\,{\rm s}$
of integration time, if the frequency and phase of the signal
at the detector are known in advance
\citep{bra98,sch99}.
The plot suggests that the prospects of detecting objects 
with $M_{\rm a} \gtrsim 10^{-5} M_{\sun}$ are encouraging,
a point first made in the context of magnetic burial by
\citet{mel00}.
Such objects include the five accreting X-ray millisecond pulsars
discovered at the time of writing, 
SAX J1808.4$-$3658, XTE J1751$-$305, XTE J0929$-$314,
XTE J1807$-$294, and XTE J1814$-$338
\citep{wij98,cha98,gal02,mar02,str03,cam03}.
Detectability is facilitated if the time-dependent Doppler shift 
from the binary orbit is known well enough to be subtracted,
but this is only practical for some objects of known $f$, 
not for an all-sky search
\citep{bra98}.

The characteristic gravitational wave strain $h_{\rm c}$
is a lower limit in an important sense: it contains the
assumption that, after averaging over polarization and orientation,
the bulk of the gravitational wave signal is emitted at $f=2\Omega$.
This is true when the observer views the system along its
rotation axis and measures
$h^+ \propto h^\times \propto e^{2i\Omega t} \sin^2\alpha$
in the two polarizations.
It is {\em not} true when the observer views the system
perpendicular to its rotation axis and measures
$h^+ \propto e^{2i\Omega t} \sin^2\alpha$
and 
$h^\times \propto e^{i\Omega t} \sin\alpha\cos\alpha$,
for example
\citep{sha83,bon96}.
One certainly expects, as a matter of chance,
to observe some systems in the latter orientation, or close to it.
For these sources, the average quantity $h_{\rm c}$
substantially underestimates the true wave strain
(which is dominated by $h^\times$ in the above example),
especially if $\alpha$ is small.\footnote{We note 
in passing that $h_{\rm c}$ also underestimates 
the true wave strain
substantially for precessing radio pulsars,
e.g.\ PSR B1828$-$11,
whose wobble angle is known to be small ($\lesssim 3^{\circ}$)
\citep{lin01}.
}
(The observed pulse modulation indices of accreting 
millisecond pulsars imply a range of $\alpha$ values.)
The effect on detectability is even greater when one takes 
into account the shape of the interferometer's noise curve;
for example, LIGO is $1.5$--$2.5$ times more sensitive at 
$\Omega \sim 0.3\,{\rm kHz}$ than at 
$2\Omega \sim 0.6\,{\rm kHz}$.

\subsection{$h_{\rm c}$ versus $\mu$
 \label{sec:gra3b}}
Polar magnetic burial is not the only mechanism whereby
accreting neutron stars with $M_{\rm a} \gtrsim 10^{-5} M_{\sun}$ 
can act as gravitational wave sources 
detectable by long baseline interferometers.
Crustal deformation due to temperature gradients
(\S\ref{sec:gra4}) is one of several alternatives
\citep{bil98,ush00}.
In principle, we can distinguish between these mechanisms
by eliminating $M_{\rm a}$ from (\ref{eq:gra6c}) and (\ref{eq:gra6d}) 
to derive a unique --- and testable --- scaling $h_{\rm c}(\mu)$ 
for magnetic burial that relates observable quantities only,
unlike (\ref{eq:gra7}), 
which features $M_{\rm a}$
(usually inferred from evolutionary models)
\citep{van95}.
The scaling,
\begin{equation}
 h_{\rm c} \propto
 (1-3/2b) [ \mu/(\psi_\ast R_\ast-\mu) + b^2/6 ]^{-1} ,
\label{eq:gra8}
\end{equation}
is graphed in Figure \ref{fig:gra3}$b$ for $3\leq b \leq 100$.
The vertical segments of all the curves represent the regime
$M_{\rm a} \ll M_{\rm c}$, where $\mu$ retains its
natal value while the ellipticity grows as $M_{\rm a}$.
The horizontal segments represent the regime
$M_{\rm a} \gg M_{\rm c}$, where $\epsilon$ saturates
while $\mu$ decreases, with the turn-off mass and
$\epsilon_{\rm max}$ scaling as $b^{-2}$.
The proportionality (\ref{eq:gra8}) does depend on $b$
and $\psi_\ast$, neither of which can be measured, 
but the relevant scalings are moderately weak
(e.g.\ $b\propto \dot{M}_{\rm a}^{-2/7} \psi_\ast^{4/7}$).
Consequently, the overall trend in Figure \ref{fig:gra3}$b$
may emerge statistically,
once many gravitational wave sources have been detected,
provided that the range of neutron star magnetic fields
at birth is relatively narrow.
Drawing upon population synthesis simulations,
\citet{har97} inferred that $\approx 90$ per cent of 
radio pulsars are born with 
$10^{23.5}\,{\rm G\,cm^2} \leq \psi_\ast \leq
 10^{24.5}\,{\rm G\,cm^2}$,
but the existence of anomalous X-ray pulsars implies
a wider range of $\psi_\ast$ in a
subset of the neutron star population
\citep{reg01}.

\section{Discussion
 \label{sec:gra4}}
\begin{figure}
\plottwo{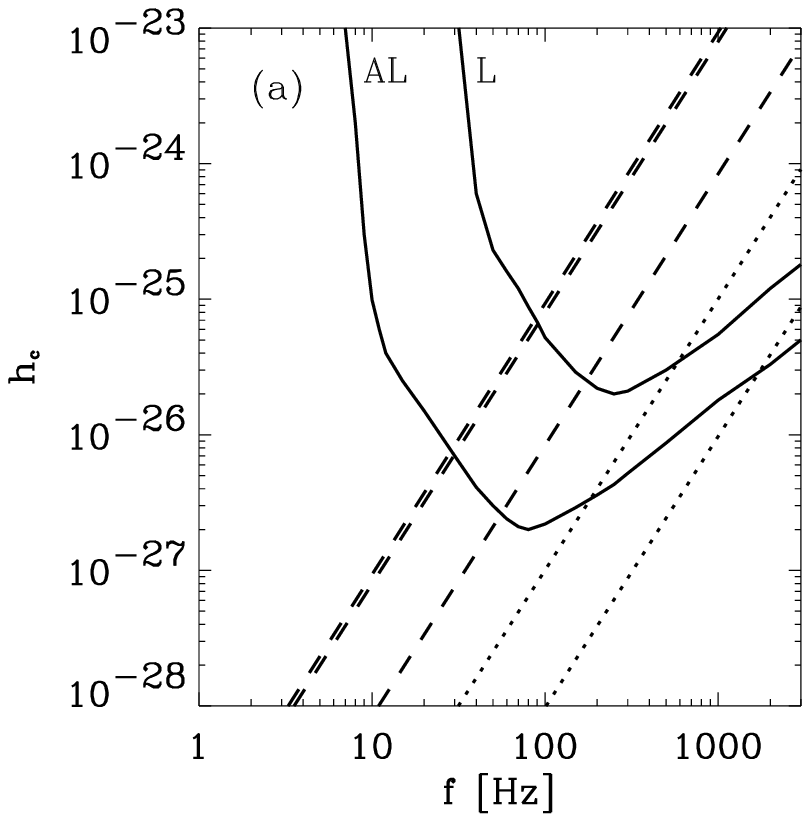}{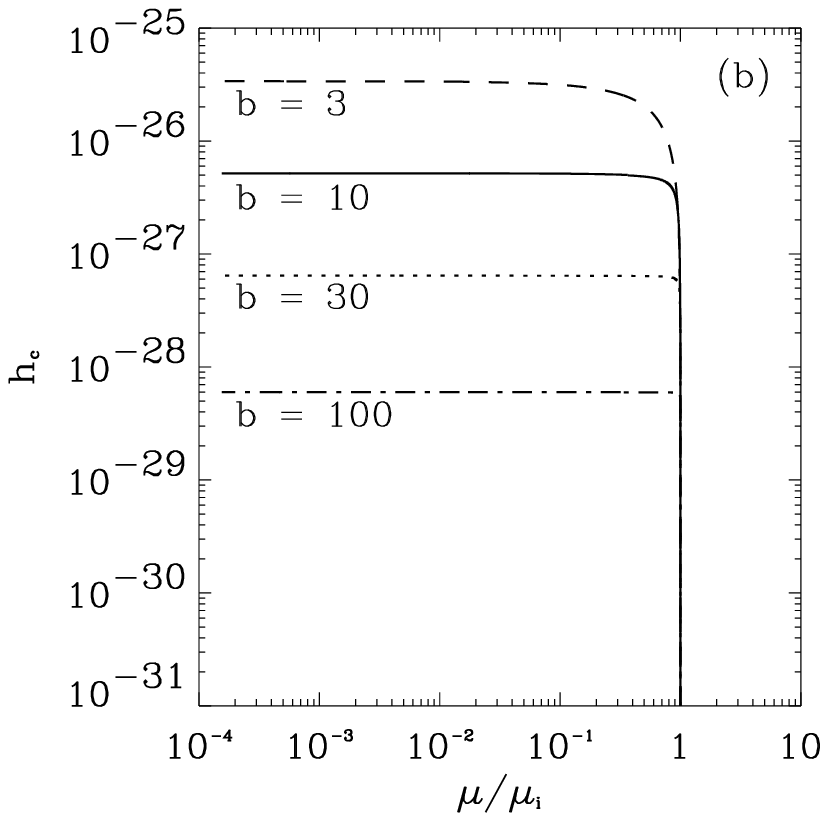}
\caption{
($a$) 
Polarization- and orientation-averaged gravitational 
wave strain $h_{\rm c}$ as a function of wave frequency $f$.
The initial and advanced LIGO noise curves (solid)
correspond to detection with 99 per cent
confidence after $10^7\,{\rm s}$ of coherent integration
\citep{sch99}.
Theoretical curves are shown for
$b=10$,
$M_{\rm a}/M_{\sun}=10^{-2}$, $10^{-4}$, $10^{-6}$
(dashed, top to bottom) 
and $b=10^2$,
$M_{\rm a}/M_{\sun}=10^{-4}$, $10^{-8}$
(dotted, top to bottom).
($b$)
Gravitational wave strain $h_{\rm c}$ versus magnetic moment $\mu$
(scaled to the natal magnetic moment $\mu_{\rm i}=\psi_\ast R_\ast$)
for $b=3$, 10, $30$, $100$,
from (\ref{eq:gra6c}) and (\ref{eq:gra6d}).
\label{fig:gra3}}
\end{figure}

In this paper, we report on two new results concerning the
gravitational radiation emitted by accreting neutron stars
undergoing polar magnetic burial.
First, upon calculating rigorously the hydromagnetic structure 
of the polar mountain,
by solving a Grad-Shafranov boundary value problem with the
correct flux freezing condition connecting the initial and
final states,
we find that the ellipticity of the star materially exceeds
previous estimates,
due to the enhanced Maxwell stress exerted by the
compressed equatorial magnetic field,
with $\epsilon \approx 20 M_{\rm c}/ M_\ast b^2$ 
for 
$M_{\rm a} \gtrsim M_{\rm c} \sim 10^{-5} M_{\sun}$,
as given by (\ref{eq:gra6e}).
The associated gravitational wave strain at Earth,
$h_{\rm c}=
 6\times 10^{-24}(M_{\rm a}/M_{\rm c})(1+M_{\rm a}b^2/8M_{\rm c})^{-1}
 (f/0.6\,{\rm kHz})^2 (d/1\,{\rm kpc})^{-1}$,
averaged over polarization and orientation,
is detectable in principle by the current generation of
long baseline interferometers, e.g.\ LIGO.
(The wave strain at $f/2=\Omega$ in one polarization
can exceed $h_{\rm c}$ for many orientations and $\alpha$ values.)
Second, the stellar magnetic moment $\mu$ decreases as the
polar mountain grows and spreads equatorward,
implying a distinctive, observable scaling $h_{\rm c}(\mu)$,
displayed in Figure \ref{fig:gra3}$b$, which can be used to test
the magnetic burial hypothesis.
Accreting neutron stars are easier to detect as
gravitational wave sources than other kilohertz
sources like coalescing neutron star binaries:
they are persistent rather than transient,
the waveform is approximately sinusoidal,
and, if X-ray pulsations or thermonuclear burst oscillations
are observed in advance (e.g.\ SAX J1808.4$-$3658),
extra sensitivity can be achieved by integrating coherently.

The range of $\epsilon$ predicted by the theory of magnetic burial
is consistent with that invoked by \citet{bil98} to explain
the clustering of spin frequencies (within 40 per cent
of $0.45\,{\rm kHz}$) of weakly magnetized, accreting neutron stars
\citep{cha03}
in terms of a stalling effect where the
gravitational wave torque ($\propto \Omega^5$) balances
the accretion torque
($\propto \dot{M}_{\rm a} R_{\rm a}^{1/2} 
  \propto \dot{M}_{\rm a}^{6/7} \mu^{2/7}$).
Importantly, the theory of magnetic burial predicts that
$\epsilon$ increases monotonically with $M_{\rm a}$ ---
a key precondition for the stalling effect to operate properly,
otherwise the stall frequency $\Omega$ 
would not be a stable fixed point.
In an alternative scenario,
\citet{bil98} and \citet{ush00} attribute 
the requisite mass quadrupole moment,
$2\times 10^{-7} I_{zz} (\dot{M}_{\rm a}/\dot{M}_{\rm Edd})^{1/2}
 (f/0.6\,{\rm kHz})^{-5/2}$,
to lateral temperature gradients in the outer crust,
which induce gradients in the electron capture rate and hence $\rho$.

Magnetic burial, acting in concert with the stalling effect,
influences the rotational evolution of an accreting neutron star
in two observationally testable ways.
First, during the late stages of accretion,
the instantaneous net torque is zero,
because the gravitational wave and accretion torques balance
\citep{bil98},
but the average torque (on the time-scale $t_{\rm a}$)
is effectively {\em negative},
because the gravitational wave torque increases with $M_{\rm a}$
as the polar mountain grows;
that is, the instantaneous stall frequency
$\Omega \propto \dot{M}_{\rm a}^{6/35} \mu(M_{\rm a})^{2/35}
 \epsilon(M_{\rm a})^{-2/5}$ decreases with $M_{\rm a}$.
This effect ought to be detectable by X-ray timing experiments 
in progress
\citep{gal02},
although it may be masked if $\dot{M}_{\rm a}$ fluctuates
stochastically on the time-scale $t_{\rm a}$,
as commonly happens.
Second, magnetic burial predicts a distinctive evolutionary
relation between $\mu$ and $\Omega$.
During the early stages of accretion,
before the star is spun up to the stall frequency,
$\Omega$ increases while $\mu$ decreases due to burial,
with 
$\Omega\propto R_{\rm a}^{1/2} M_{\rm a}
 \propto \dot{M}_{\rm a}^{-1/7}  \mu^{-5/7}$
in the regime $\mu\propto M_{\rm a}^{-1}$.
However, during the late stages of accretion,
$\Omega$ and $\mu$ both decrease as explained above.
Hence there exists a maximum spin frequency
$\Omega/2\pi \lesssim 1\,{\rm kHz}$
bounding the population of accreting millisecond pulsars
in the $\Omega$-$\mu$ plane, and objects follow
$\Lambda$-shaped evolutionary tracks in that plane.

Once accretion ends, do we expect to detect the neutron star
as a radio millisecond pulsar?
There are two arguments against this from the perspective
of magnetic burial.
First, the gravitational wave spin-down time,
$t_{\rm g}\approx 5\times 10^7 (\epsilon/10^{-7})^{-2}
 (\Omega/2\,{\rm krad\,s^{-1}})^{-4}\,{\rm yr}$,
is shorter than the observed age,
so the neutron star rapidly brakes
below the radio pulsar death line 
($\Omega/2\pi < 0.1\,{\rm kHz}$)
and is extinguished as a radio source.
Note that this occurs no matter what generates the quadrupole
moment inferred from the stalling effect.
Second, this rapid spin-down is interrupted if the
stellar deformation relaxes quickly (compared to $t_{\rm g}$),
for example if the buried, polar magnetic field is resurrected
on the Ohmic time-scale $t_{\rm d}$ of the outer crust, 
which satisfies 
$t_{\rm d} \sim t_{\rm a}
 = 6\times 10^7 (M_{\rm a}/M_{\sun})
  (\dot{M}_{\rm a}/\dot{M}_{\rm Edd})^{-1}
 \,{\rm yr}$
in the regime $M_{\rm a}\gtrsim 10^{-5}M_{\sun}$.
However, the buried field is resurrected in stages;
the steepest gradients are smoothed out ($L\propto t_{\rm a}^{1/2}$),
but the natal field is not fully restored over the typical
lifetime of a millisecond pulsar, 
because one has
$\mu \propto L\propto t_{\rm a}^{1/2}$ and hence
$B \sim 10^{9-10}\,{\rm G}$ after $\sim 10^9\,{\rm yr}$
\citep{cum01,mel01}.
This scenario leaves the radio millisecond pulsars with the
lowest fields unexplained unless they have low fields initially.


\acknowledgments
The authors are grateful to Sterl Phinney for many
illuminating discussions.
This research was supported in part by 
Australian Research Council grant DP--0208735,
NASA Grants NAG5--2756 and NAG5--3073,
NSF Grant AST--95--28271,
the Miller Institute for Basic Research in Science
through a Miller Fellowship,
and by an Australian Postgraduate Award.

\bibliographystyle{apj}
\bibliography{accretion}

\begin{thebibliography}{43}
\expandafter\ifx\csname natexlab\endcsname\relax\def\natexlab#1{#1}\fi

\bibitem[{{Andersson}(1998)}]{and98}
{Andersson}, N. 1998, \apj, 502, 708

\bibitem[{{Bildsten}(1998)}]{bil98}
{Bildsten}, L. 1998, \apjl, 501, L89+

\bibitem[{{Bonazzola} \& {Gourgoulhon}(1996)}]{bon96}
{Bonazzola}, S., \& {Gourgoulhon}, E. 1996, \aap, 312, 675

\bibitem[{{Brady} \& {Creighton}(2000)}]{bra00}
{Brady}, P.~R., \& {Creighton}, T. 2000, \prd, 61, 82001

\bibitem[{{Brady} {et~al.}(1998){Brady}, {Creighton}, {Cutler}, \&
  {Schutz}}]{bra98}
{Brady}, P.~R., {Creighton}, T., {Cutler}, C., \& {Schutz}, B.~F. 1998, \prd,
  57, 2101

\bibitem[{{Brown} \& {Bildsten}(1998)}]{bro98}
{Brown}, E.~F., \& {Bildsten}, L. 1998, \apj, 496, 915

\bibitem[{{Campana} {et~al.}(2003){Campana}, {Ravasio}, {Israel}, {Mangano}, \&
  {Belloni}}]{cam03}
{Campana}, S., {Ravasio}, M., {Israel}, G.~L., {Mangano}, V., \& {Belloni}, T.
  2003, \apjl, 594, L39

\bibitem[{{Chakrabarty} \& {Morgan}(1998)}]{cha98}
{Chakrabarty}, D., \& {Morgan}, E.~H. 1998, \nat, 394, 346

\bibitem[{{Chakrabarty} {et~al.}(2003){Chakrabarty}, {Morgan}, {Muno},
  {Galloway}, {Wijnands}, {van der Klis}, \& {Markwardt}}]{cha03}
{Chakrabarty}, D., {Morgan}, E.~H., {Muno}, M.~P., {Galloway}, D.~K.,
  {Wijnands}, R., {van der Klis}, M., \& {Markwardt}, C.~B. 2003, \nat, 424, 42

\bibitem[{{Cumming}(2002)}]{cum02}
{Cumming}, A. 2002, \mnras, 333, 589

\bibitem[{{Cumming} {et~al.}(2004){Cumming}, {Arras}, \& {Zweibel}}]{cum04}
{Cumming}, A., {Arras}, P., \& {Zweibel}, E. 2004, \apj, 609, 999

\bibitem[{{Cumming} {et~al.}(2001){Cumming}, {Zweibel}, \& {Bildsten}}]{cum01}
{Cumming}, A., {Zweibel}, E., \& {Bildsten}, L. 2001, \apj, 557, 958

\bibitem[{{Cutler}(2002)}]{cut02}
{Cutler}, C. 2002, \prd, 66, 84025

\bibitem[{{Dhurandhar} {et~al.}(1996){Dhurandhar}, {Blair}, \& {Costa}}]{dhu96}
{Dhurandhar}, S.~V., {Blair}, D.~G., \& {Costa}, M.~E. 1996, \aap, 311, 1043

\bibitem[{{Galloway} {et~al.}(2002){Galloway}, {Chakrabarty}, {Morgan}, \&
  {Remillard}}]{gal02}
{Galloway}, D.~K., {Chakrabarty}, D., {Morgan}, E.~H., \& {Remillard}, R.~A.
  2002, \apjl, 576, L137

\bibitem[{{Geppert} \& {Urpin}(1994)}]{gep94}
{Geppert}, U., \& {Urpin}, V. 1994, \mnras, 271, 490

\bibitem[{{Hameury} {et~al.}(1983){Hameury}, {Bonazzola}, {Heyvaerts}, \&
  {Lasota}}]{ham83}
{Hameury}, J.~M., {Bonazzola}, S., {Heyvaerts}, J., \& {Lasota}, J.~P. 1983,
  \aap, 128, 369

\bibitem[{{Hartman} {et~al.}(1997){Hartman}, {Bhattacharya}, {Wijers}, \&
  {Verbunt}}]{har97}
{Hartman}, J.~W., {Bhattacharya}, D., {Wijers}, R., \& {Verbunt}, F. 1997,
  \aap, 322, 477

\bibitem[{{Heyl}(2000)}]{hey00}
{Heyl}, J.~S. 2000, \mnras, 317, 310

\bibitem[{{Ioka}(2001)}]{iok01}
{Ioka}, K. 2001, \mnras, 327, 639

\bibitem[{{Jones}(1975)}]{jon75}
{Jones}, P.~B. 1975, \apss, 33, 215

\bibitem[{{Katz}(1989)}]{kat89}
{Katz}, J.~I. 1989, \mnras, 239, 751

\bibitem[{{Konno} {et~al.}(2000){Konno}, {Obata}, \& {Kojima}}]{kon00}
{Konno}, K., {Obata}, T., \& {Kojima}, Y. 2000, \aap, 356, 234

\bibitem[{{Lindblom} {et~al.}(1998){Lindblom}, {Owen}, \& {Morsink}}]{lin98}
{Lindblom}, L., {Owen}, B.~J., \& {Morsink}, S.~M. 1998, Physical Review
  Letters, 80, 4843

\bibitem[{{Link} \& {Epstein}(2001)}]{lin01}
{Link}, B., \& {Epstein}, R.~I. 2001, \apj, 556, 392

\bibitem[{{Litwin} {et~al.}(2001){Litwin}, {Brown}, \& {Rosner}}]{lit01}
{Litwin}, C., {Brown}, E.~F., \& {Rosner}, R. 2001, \apj, 553, 788

\bibitem[{{Markwardt} {et~al.}(2002){Markwardt}, {Swank}, {Strohmayer}, {Zand},
  \& {Marshall}}]{mar02}
{Markwardt}, C.~B., {Swank}, J.~H., {Strohmayer}, T.~E., {Zand}, J.~J.~M.~i.,
  \& {Marshall}, F.~E. 2002, \apjl, 575, L21

\bibitem[{{Melatos} \& {Phinney}(2000)}]{mel00}
{Melatos}, A., \& {Phinney}, E.~S. 2000, in ASP Conf. Ser. 202: IAU Colloq.
  177: Pulsar Astronomy - 2000 and Beyond, 651--+

\bibitem[{{Melatos} \& {Phinney}(2001)}]{mel01}
{Melatos}, A., \& {Phinney}, E.~S. 2001, Publications of the Astronomical
  Society of Australia, 18, 421

\bibitem[{{Mouschovias}(1974)}]{mou74}
{Mouschovias}, T.~C. 1974, \apj, 192, 37

\bibitem[{{Palomba}(2001)}]{pal01}
{Palomba}, C. 2001, \aap, 367, 525

\bibitem[{{Payne} \& {Melatos}(2004)}]{pay04}
{Payne}, D.~J.~B., \& {Melatos}, A. 2004, \mnras, 351, 569

\bibitem[{{Phinney}(1991)}]{phi91}
{Phinney}, E.~S. 1991, \apjl, 380, L17

\bibitem[{{Regimbau} \& {de Freitas Pacheco}(2001)}]{reg01}
{Regimbau}, T., \& {de Freitas Pacheco}, J.~A. 2001, \aap, 374, 182

\bibitem[{{Schutz}(1999)}]{sch99}
{Schutz}, B.~F. 1999, Classical and Quantum Gravity, 16, A131

\bibitem[{{Shapiro} \& {Teukolsky}(1983)}]{sha83}
{Shapiro}, S.~L., \& {Teukolsky}, S.~A. 1983, {Black holes, white dwarfs, and
  neutron stars: The physics of compact objects} (Research supported by the
  National Science Foundation.~New York, Wiley-Interscience, 1983, 663 p.)

\bibitem[{{Shibazaki} {et~al.}(1989){Shibazaki}, {Murakami}, {Shaham}, \&
  {Nomoto}}]{shi89}
{Shibazaki}, N., {Murakami}, T., {Shaham}, J., \& {Nomoto}, K. 1989, \nat, 342,
  656

\bibitem[{{Strohmayer} {et~al.}(2003){Strohmayer}, {Markwardt}, {Swank}, \&
  {in't Zand}}]{str03}
{Strohmayer}, T.~E., {Markwardt}, C.~B., {Swank}, J.~H., \& {in't Zand}, J.
  2003, \apjl, 596, L67

\bibitem[{{Taam} \& {van de Heuvel}(1986)}]{taa86}
{Taam}, R.~E., \& {van de Heuvel}, E.~P.~J. 1986, \apj, 305, 235

\bibitem[{{Uchida} \& {Low}(1981)}]{uch81}
{Uchida}, Y., \& {Low}, B.~C. 1981, Journal of Astrophysics and Astronomy, 2,
  405

\bibitem[{{Ushomirsky} {et~al.}(2000){Ushomirsky}, {Cutler}, \&
  {Bildsten}}]{ush00}
{Ushomirsky}, G., {Cutler}, C., \& {Bildsten}, L. 2000, \mnras, 319, 902

\bibitem[{{van den Heuvel} \& {Bitzaraki}(1995)}]{van95}
{van den Heuvel}, E.~P.~J., \& {Bitzaraki}, O. 1995, \aap, 297, L41+

\bibitem[{{Wijnands} \& {van der Klis}(1998)}]{wij98}
{Wijnands}, R., \& {van der Klis}, M. 1998, \nat, 394, 344

\end{thebibliography}

\clearpage



\end{document}